# Magnetic response of mesoscopic rings: a quantum size effect

J. W. Ding[*,1,3], X. H. Yan[1], B. G. Wang[2], D. Y. Xing[2]

[1] *College of Electronic Science Engineering, Nanjing University of Post & Telecommunication, 210046, Nanjing, China*

[2] *National Laboratory of Solid State Microstructures, Nanjing University, Nanjing 210093, China*

[3] *Department of Physics & Institute for Nanophysics and Rare-earth Luminescence, Xiangtan University, Xiangtan, 411105, Hunan, China*

**Abstract:** We analytically study the magnetic response of persistent current (PC) in normally non-interacting mesoscopic rings of bimodal potential with nearest neighboring interactions (*t*) and alternating site energies. It is shown that a ring of perimeter (*N*) and width (*M*) generally shows weak diamagnetic, breaking the even-odd rule of electron filling. Especially, a maximal paramagnetic current in primary $\Phi_0/2$ period is predicted at $N=(2p+1)(M+1)$ with odd *M* and integer *p*, while a maximal diamagnetic $\Phi_0/2$- current obtained at $N=(2p+1)(M+1)\pm 1$ with even *M*. The current amplitudes depend strongly on both *N* and *M*, varied by at least 1~2 orders of magnitude, exhibiting a remarkable quantum size effect. A current limit of paramagnetic harmonics $A_l=8t/lp\pi\Phi_0$ is expected at $N=2p(M+1)$ with flux quantum $\Phi_0=h/e$, independent of the sizes of *N* and *M*, in favor of experiment observation. A new mechanism of magnetic response is proposed that an electron circling the ring shall pass successively each channel within one flux quantum, accumulating an additional phase on each inter-channel transition, which leads to the paramagnetic-diamagnetic transition and period halving. The results unify and unveil the contradictions in PC between theory and experiments, validating quantum mechanics at mesoscopic scale.

Quantum mechanics predicts a persistent current[1,2] (PC) in normal mesoscropic ring threaded by magnetic flux $\Phi$, in period of flux quantum $\Phi_0=h/e$, which flows forever without dissipating energy. This seemingly preposterous effect had been observed in several experiments. The first evidence for the existence of PC was reported by Levy *et al.*[3], and a series of experiments were subsequently made on metallic and semiconducting rings[4-12]. However, the sign, amplitude and period halving observed in PC are in apparent contradiction with theory and even among the experiments themselves, challenging in return the validity of quantum mechanics at mesoscopic scale.

Experimentally, the measured currents in several experiments[3,9,10] were larger by at least 2 orders of magnitude than theoretical prediction[1,2,13-15], while the currents in other several experiments[5,8,11,12] had an overall magnitude in agreement with theory. For the periodicity and direction of PC, a *paramagnetic* $\Phi_0$-current was observed in single Au ring[4], while the currents in Au rings array exhibit a *diamagnetism* in $\Phi_0$- and $\Phi_0/2$-periods[9]. From the experiments of collective rings[3,6], it was found that *only even harmonics* ($\Phi_0/2$-current) does not average to zero, showing *consistently diamagnetic*. In Bluhm's experiment[11], furthermore, it was *in situ* measured that the *sign and amplitude* of PC vary between nominally identical rings. Especially, recent experiment[12] had confirmed *diffusive non-interacting electrons* in normal metal rings, arousing the requirement of a simple and effective PC theory.

Theoretically, it was predicted that PC has *a priori* a random sign in single one-dimensional (1D) ring, depending on the *parity* of electron occupation ($N_e$) and the actual disorder[13-15]. To have a $\Phi_0/2$ period, it was conjectured that a wave packet must enclose the flux twice[3]. From some numerical calculations[14-17], the $\Phi_0/2$-current had been obtained, however, it presents *barely exclusively paramagnetic*, in contradiction to the experiments[3,6]. Much more theoretical works further focused on some subtle effects[18-24] on PC such as *e-e* interactions and spin-orbit scattering. While theories are more and more complicated, the magnetic response of PC remains a topic of controversy in mesoscopic condensed-matter physics[25].

Here we study the magnetic response of PC in normally non-interacting mesoscopic rings of perimeter $L=Na$ and width $W=Ma$ with lattice constant $a$ and unit cell number $N_c=N/2$, as shown in Fig.1. The non-interacting Hamiltonian of bimodal potential is $H=\sum\varepsilon_i|i\rangle\langle i|+\sum t_{ij}|i\rangle\langle j|$, with nearest neighboring interactions ($t_{ij}=t$) and alternating site energies ($\varepsilon_i=\varepsilon_1$ and $\varepsilon_2$). In the presence of $\Phi$, $t_{ij}$ along $L$ direction are modified by a phase factor $t_{ij}=t\exp(2\pi i\phi/N)$ with $\phi=\Phi/\Phi_0$. Such model ring of $\varepsilon_1\neq\varepsilon_2$ can well depict the main characteristics of both a semiconducting ring with an energy gap and a normal metallic ring with a disorder of particular configuration.

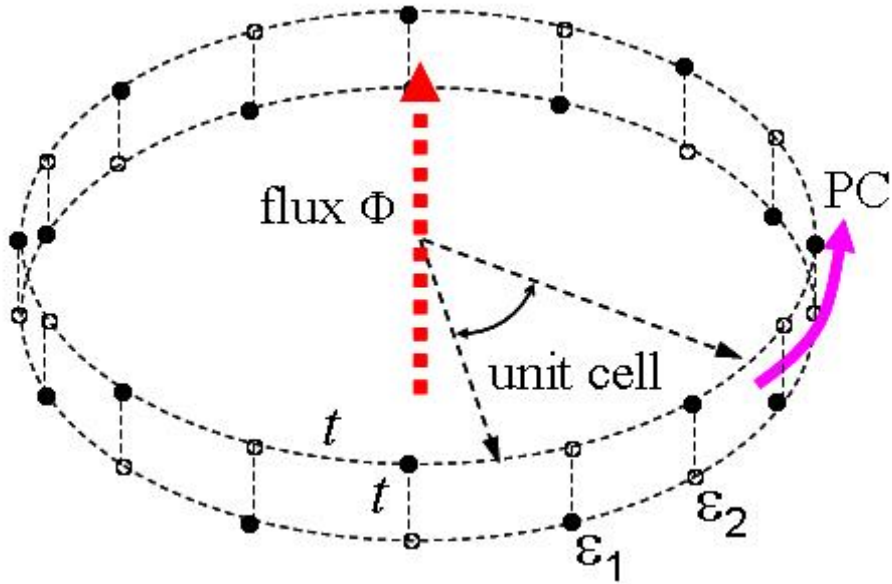

Figure 1 (Color online) Schematic of a mesoscopic ring of $N$=16 and $M$=2 with alternating site energies $\varepsilon_1$ and $\varepsilon_2$, threaded by a magnetic flux $\Phi$.

No loss of generality, we take $\varepsilon_1=-\varepsilon_2=\varepsilon\geq 0$. The energy $E_n$ and current $i_n$ are analytically derived by developing a super-cell method[26-28],

$$E_n = \pm\sqrt{\varepsilon^2 + 4t^2[\cos((n+\phi)\pi/N_c) + \Delta_j]^2}, \qquad (1)$$

$$i_n = \frac{\pi I_0}{NE_n}[\sin\frac{2\pi(n+\phi)}{N_c} + 2\Delta_j \sin\frac{\pi(n+\phi)}{N_c}], \qquad (2)$$

where $I_0=4t/\Phi_0$ and $\Delta_j=\cos[j\pi/(M+1)]$ with integers $n$ and $j$ ($0\leq n\leq N_c-1$, $1\leq j\leq M$). Total current is obtained by $I=\sum i_n$ at zero temperature ($T=0$) and $I=\sum i_n f_n$ at finite $T$ with $f_n$ Fermi distribution, of which Fourier spectrum $A_l$ is derived from $A_l=2\int d\phi I\sin 2\pi l\phi$ with integrating range [-0.5, 0.5].

In Fig.2 we show the dispersions and currents of some specific rings and the Fourier spectra in Fig.3. Due to $\varepsilon\neq 0$, the dispersion is split into a valence band (VB) and a conducting band (CB) with a gap $E_g\sim 2\varepsilon$, exhibiting a hole-electronic symmetry [Fig.2(a-d)]. A little $\varepsilon$ induces dramatic changes in the highest VB (HVB) such as energy gap and level anti-crossings, a series of energy maxima existing there. Such special energy spectra imply some nontrivial properties of PC.

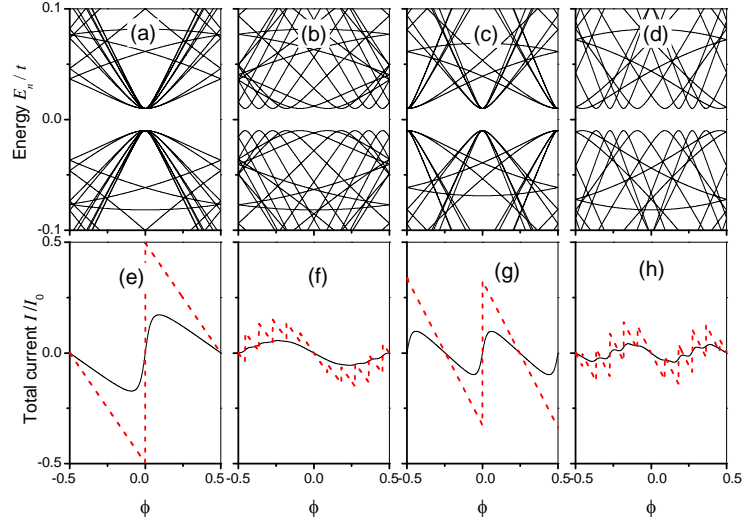

Figure 2 (Color online) Dispersions and currents of $N_c=p(M+1)+q$ rings with $\varepsilon_1=-\varepsilon_2=\varepsilon=0.01t$. (a) $M=10$, $p=2$, $q=0$; (b) $M=10$, $p=2$, $q=1$; (c) $M=11$, $p=1$, $q=6$; (d) $M=10$, $p=1$, $q=5$. The currents are correspondingly shown in (e)-(h) under the occupation of $N_e=MN_c$ at zero temperature ($T=0$) for $\varepsilon=0$ (dashed line) and $\varepsilon=0.01t$ (solid line).

From Fig.2(e-h), the current in perfect ring exhibits a series of step discontinuities, which are smoothen into a sinusoidal form with $\varepsilon\neq 0$, no step existing at zero flux limit. A paramagnetic or diamagnetic current in primary $\Phi_0$- or $\Phi_0/2$-period is distinctly observed in those specific rings. From Fig.3, only the first two lowest harmonics may survive at finite temperature, while higher harmonics are predicted in perfect rings at $T=0$. Hereafter one can focus on the two lowest harmonics.

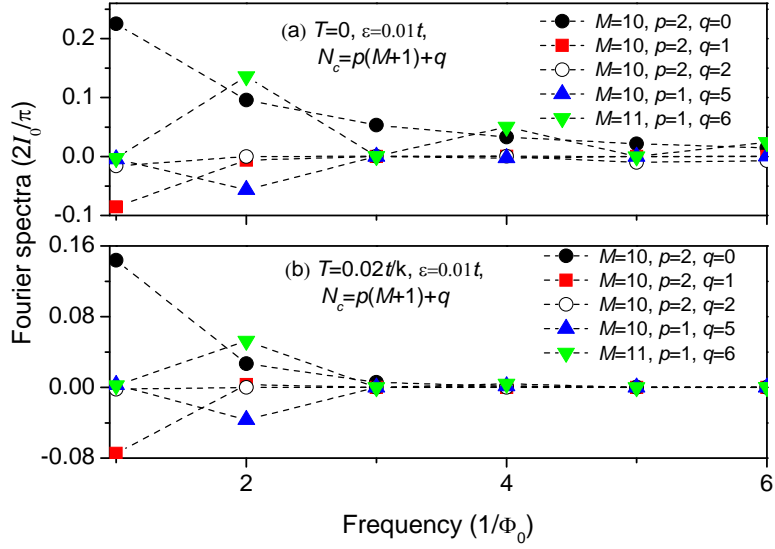

Figure 3 (Color online) Fourier spectra of PC in some specific rings at (a) $T=0$ and (b) $T=0.02t/k$ with k Boltzmann constant.

For a general ring of $N_c=p(M+1)+q$, Fig.4 shows Fourier ampere $A_1$ and $A_2$ as a function of ring perimeter ($N_c$) under the electron occupation of $N_e=MN_c$. Clearly, both amplitudes and signs of $A_1$ and $A_2$ are dramatically modulated only by changing $N_c$ with $q=0 \sim M$, which have a variation of at least 1~2 orders of magnitude, exhibiting a remarkable quantum size effect. This well explains the serious discrepancy of the current amplitude between theory and experiments[3-15].

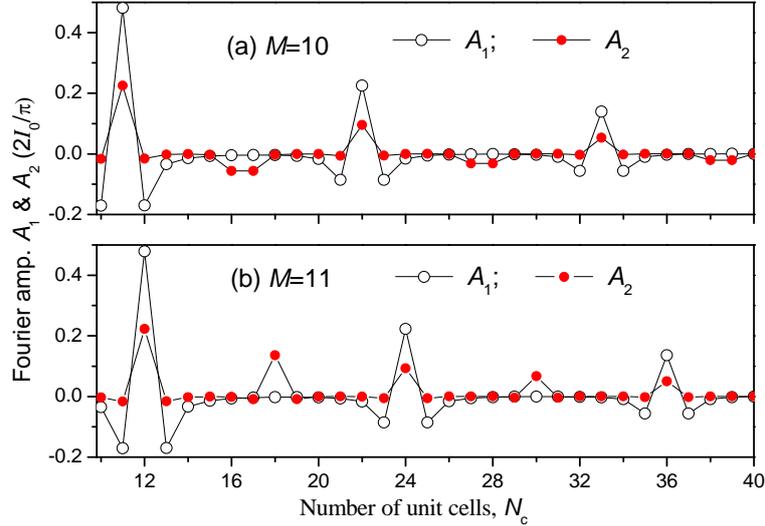

Figure 4 (Color online) Fourier ampere $A_1$ and $A_2$ of PC in some specific rings as a function of $N_c$ the number of unit cells in the case of $\varepsilon=0.01t$ and $T=0$ for (a) even $M$ and (b) odd $M$.

For the sign of PC, in general, a ring shows weak diamagnetic, irrespective of the parity of $N_e$, breaking the even-odd rule of electron filling. Whether even or odd $M$, interestingly, a maximal *paramagnetism* ($A_l$>0) appears at $N_c=p(M+1)$ with $q=0$, while a maximal *diamagnetism* ($A_l$<0) at $N_c=p(M+1)\pm1$ with $q=\pm1$. This well explains the experiments of *paramagnetic*[4], *diamagnetic* response[9] of both $\Phi_0$- and $\Phi_0/2$-currents. Only by adding or subtracting a unit cell in $N_c$, surprisingly, there occurs a *paramagnetic-diamagnetic* transition, consistent with the experiment of nominally identical rings[11].

In the case of odd *M*, furthermore, a maximal *paramagnetic* $\Phi_0/2$-current ($A_2$>0) appears only at $N_c=(p+1/2)(M+1)$ with $q=(M+1)/2$, while its $\Phi_0$-current is *weak diamagnetic* ($A_1$<0). Such a *paramagnetic* $\Phi_0/2$-current had not been experimentally reported so far, due to a sharp geometrical condition. In the case of even *M*, especially, a maximal *diamagnetic* $\Phi_0/2$-current ($A_2$<0) exists at $q=M/2$ and $M/2+1$, while its *weak* $\Phi_0$-current is *diamagnetic* ($A_1$<0). This result is consistent with the experimental observation on the *consistent diamagnetism* in $\Phi_0/2$ period[3,6].

For a real assemblage of nominally identical rings, the observability of $\Phi_0/2$-current depends on the ensemble average. In the presence of a fluctuation in perimeter ($N_c$) with mean value $N_0$ and standard deviation $\sigma$, Fig.5 shows the averaged currents over the sample number $N_s$ at even *M*. Except for finite fluctuations, the *diamagnetic* $\Phi_0/2$-current survives in the average, while the *diamagnetic* $\Phi_0$-current is very weak. The results differ from the ensemble average over electron number and/or over chemical potential[16,17], a *paramagnetic* $\Phi_0/2$-current appearing in the former and *no period halving* in the latter. This shows that the *period halving* originates only from the quantum size effect, with no requirement of particle spin[18].

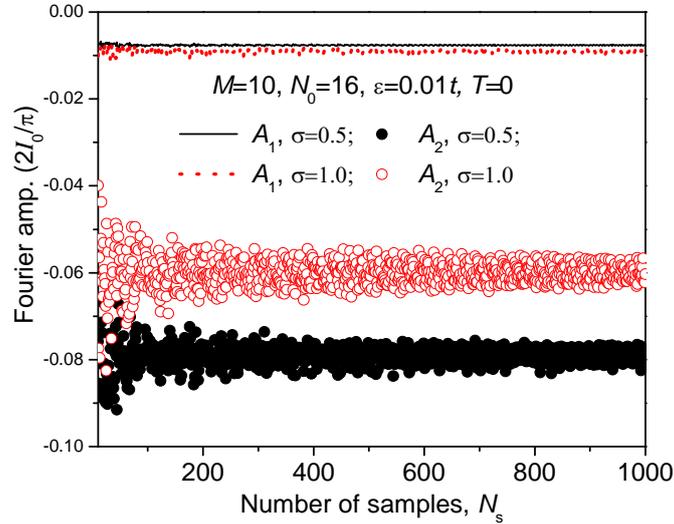

Figure 5 (Color online) The averaging Fourier ampere over $N_s$ the sample number of an even *M* rings in the presence of a fluctuation in $N_c$ on mean value $N_0$=16, *M*=10, and standard deviation $\sigma$ at *T*=0 and $\varepsilon$=0.01*t*.

To understand the quantum size effect in magnetic response, the currents and Fourier spectra are further derived analytically from perfect rings of finite width. From Eqs.(1) and (2), the locations of the energy maxima of HVB in channel *j* at

$\phi_j=\phi_\pm$ are determined by[29]

$$n_j + \phi_j = N_c \pm j[p + q/(M+1)], \qquad (3)$$

with $-1/2 \leq \phi_j \leq 1/2$ and $n_j \in n$ or $n\pm 1/2$, which depend sensitively on ring geometry ($N_c$ and $M$). For another channel $j'=M+1-j$, similarly, its energy maxima are located at the same position $\phi=\phi_j$ as in channel $j$, forming an intersecting parabola. For lower free energy, naturally, an electron circling the ring shall transfer from one channel to another at $\phi=\phi_j$ and pass successively each channel within one flux quantum, indicating a new mechanism of magnetic response.

In channel $j$, the electron occupation $N_j$ can be obtained from Eq.(3) by $n_j$, of which the Fourier contribution $A_{j,V}$ is derived by[29]

$$A_{j,V} = [\nu + (-1)^l(1-\nu)]\frac{2I_0}{lN}\sin\frac{N_j\pi}{N}, \qquad (4)$$

with $\nu=\mathrm{mod}(N_j-1,2)$. From Eq. (4), no *diamagnetic* even harmonic ($l=2$) in single channel is in contradiction to the experiment of diamagnetic $\Phi_0/2$-current, showing that 1D ring model is oversimplified. Importantly, it is found the magnetic response depends the electron occupation $N_j$, a maximum existing at half filling, exhibiting a shell-like structure of *d*-electrons. Such a non-monotonous behavior was neglected in previous theory, differing essentially from the linear raise in free electron model, from which some new characteristic of magnetic response can be expected.

The joint Fourier contribution $A_{j,j'}$ of the dual channel currents $I_j$ and $I_{j'}$ is separately derived in terms of subsection integral by[29]

$$A_{j,j'} = \frac{4I_0}{lN}\cos(2l\pi\phi_j)\sin\frac{j\pi}{M+1}, \qquad (5)$$

depending sensitively on both $l$ and $\phi_j$. Obviously, the electron accumulates an additional phase of $\cos(2\pi l\phi_j)$ on each inter-channel transition, having an contribution of different importance to total Fourier spectra $A_l$, $A_l=\sum A_{j,j'}$. This unveils a new physics picture of electron transport in mesoscopic rings.

At $N_c=p(M+1)$ with $q=0$, the channel currents exhibit a perfect phase correlation, $\phi_j=0$ for any $j$. Both even and odd harmonics of $A_l=2I_0/\pi pl$ are derived at large $M$, assuming a *maximal paramagnetism*. Among all ring geometries, especially, A current limit of $A_l=8t/l\pi\Phi_0$ is expected at $p=1$, independent of the sizes of $N$ and $M$, in favor of experimental observation.

At $N_c=p(M+1)\pm 1$ with $q=\pm 1$, $\phi_j=\pm j/(M+1)$ for $j\leq(M+1)/2$ and $j'=M+1-j$. Both even and odd harmonics are given by $A_l=-2I_0/\pi pl(4l^2-1)$ at large $M$, showing a *maximal diamagnetism*. The phase accumulation induces a sign-mutation and thus a *paramagnetic-diamagnetic transition* as observed in experiments[4,9,11]. Also, harmonic decays rapidly by $\sim 1/l(4l^2-1)$, less by about 1/3 and 1/15 at $l=1$ and 2 than that in previous theory. This helps explaining the fact that higher harmonics had hardly been observed.

At $N_c=(p+1/2)(M+1)$ with odd $M$, all the channels are grouped into two groups, $\phi_j=0$ for even $j$ and $\phi_j=\pm 1/2$ for odd $j$, between which there appears an additional phase difference $l\pi$. Due to coherent phase correlation, even harmonic exhibits a *maximal paramagnetism*, $A_l=4I_0/\pi l(2p+1)$ at large $M$, while odd harmonic presents *weak diamagnetic* by destructive phase correlation, $A_l=-\pi I_0/lN(M+1)$. The ratio of $A_2$ to $A_1$ is about $-2(M+1)^2/\pi^2$, at least 1~2 orders of magnitude at $M=11$~21. Therefore, the current exhibits a *paramagnetism* in primary $\Phi_0/2$ period, with no requirement of one electron enclosing the flux twice[3].

At $N_c=(p+1/2)(M+1)\pm1/2$ with even $M$, $\phi_j=\pm[1/2-j/2(M+1)]$ for odd $j$, while $\phi_j=\pm j/2(M+1)$ for even $j$. Even harmonic of $A_l=-2I_0/\pi(2p+1)l(l^2-1)$ is derived at large $M$, showing a *maximal diamagnetism*, while odd harmonic presents *weak diamagnetic*, $A_l=-\pi I_0/lN(M+1)$. Also, higher even harmonic decays rapidly by $\sim 1/l(l^2-1)$, and thus the current shows a maximal *diamagnetism* in primary $\Phi_0/2$ period. This result well explains the tricky experiment of the *consistent diamagnetic* $\Phi_0/2$-current[3,6]..

In conclusion, we have analytically studied the magnetic response of a non-interacting mesoscopic ring of bimodal potential. The contradictions in PC between theory and experiments are well unified and unveiled, validating quantum mechanics at mesoscopic scale. Also, our results confirmed the experimental observation of non-interacting electrons, which can be further applied to explore more complex phenomena in newfashioned micro-nano structures. Furthermore, a series of new findings and/or predictions such as the current limit, period halving, and shell-like structure of magnetic response may be correlated with the problems of Aharonov-Bohm and universal conductance fluctuations of *finite amplitude* in nanowires and mesoscopic rings, which may help build a universal physics picture of electron transport in mesoscopic systems.